\documentclass{emulateapj}

\def\hi{\ifmmode {\rm H}\,{\sc i}~ \else H\,{\sc i}~\fi}

\def\ovii{O\,{\sc vii}}

\def\chandra {\emph{Chandra}}

\slugcomment{Accepted to ApJ Letters}

\shorttitle{Hosts of X-ray absorption systems}
\shortauthors{Williams, Mulchaey, \& Kollmeier}

\begin{document}

\title{Warm-hot gas in groups and galaxies toward H2356-309$^\star$}

\author{Rik J. Williams, John S. Mulchaey, \& Juna A. Kollmeier}
\affil{Carnegie Observatories, 813 Santa Barbara St., Pasadena, 
CA 91101, USA}
\altaffiltext{$^\star$}{This paper includes data gathered with the 6.5 meter Magellan 
Telescopes located at Las Campanas Observatory, Chile.}
\email{williams@obs.carnegiescience.edu}

\begin{abstract}
We present a detailed analysis of the galaxy and group distributions
around three reported X-ray absorption line systems in the spectrum
of the quasar H2356-309. Previous studies associated these absorbers 
with known large-scale galaxy structures (i.e., walls and filaments) along
the line of sight.  Such absorption lines typically trace $\sim 10^{5-7}$\,K gas, 
and may be evidence of the elusive warm-hot
intergalactic medium (WHIM) thought to harbor the bulk of the low-redshift
``missing baryons;'' alternatively, they may be linked to individual
galaxies or groups in the filaments.  Here we combine
existing galaxy survey data with new, multi-object \emph{Magellan} spectroscopy 
to investigate the detailed galaxy distribution near each absorber.  All of these three
absorption systems are within the projected virial radii of nearby galaxies
and/or groups, and could therefore arise in these virialized structures
rather than (or in addition to) the WHIM.  However, we find no additional galaxies near
a fourth ``void'' absorber recently found in the spectrum,
suggesting that this system may indeed trace gas unassociated with any
individual halo.
Though the number of known systems is still small, spatial coincidences suggest
that some X-ray absorbers lie in galaxy and/or group environments, though others
could still trace the large-scale filamentary WHIM gas predicted by
simulations.
\end{abstract}


\section{Introduction}
Few phenomena in modern astrophysics exhibit the maddening combination
of theoretical ubiquity and observational elusiveness as the warm-hot
intergalactic medium (WHIM).  Although it is effectively a generic
prediction of hydrodynamic simulations, containing up to 50\% of
the baryons in the low-redshift universe \citep{dave10}, decisive
observational confirmation of its existence has yet to be found.
At high redshifts, the IGM has long been assayed through measurements
of the Lyman-alpha forest.  At $z\la 2$, however, this forest progressively
thins to a desert: ultraviolet spectra of bright quasars 
reveal drastically fewer absorbers resulting in an inferred local 
baryon density far below what is seen at high redshift.

This discrepancy is often described as a problem of ``missing baryons.''
Of course, the matter traced by the high-redshift Ly$\alpha$ forest has not vanished, but
instead has become ionized (through shock heating 
and photoionization from the UV/X-ray background) to the point
where essentially no neutral hydrogen is left.  Fortunately, at typical WHIM
temperatures and densities ($T=10^5-10^7$\,K and $n=10^{-6}-10^{-4}$\,cm$^{-3}$,
respectively), species of common metals like carbon, oxygen, and neon
retain a few electrons, in principle allowing UV and X-ray absorption line 
studies of the WHIM at $z<2$ \citep{perna98,cen99}.  Such measurements are challenging (particularly at X-ray
wavelengths); nonetheless, several X-ray forest lines have been reported
in blind searches with ultra-deep \chandra\ and \emph{XMM} 
spectroscopy \citep{fang02,fang07,nicastro05a,nicastro05b,williams07}.

While blind searches provide an unbiased census of
the UV/X-ray absorber population, \emph{targeted} observations
reveal important information about the nature of the systems themselves.
\citet{buote09} successfully employed this tactic with deep \chandra\
spectroscopy of the X-ray bright $z=0.165$ 
quasar H$2356-309$ (H2356 hereafter), a system which has proven to be a rich laboratory
for connecting putative IGM lines to large-scale galaxy structures.
Not only did they detect an O\,{\sc vii} absorption
line (corresponding to $\sim 10^6$\,K gas) at the redshift of the 
Sculptor Wall supercluster ($z=0.03$), but
follow-up studies on the same sightline by \citet{zappacosta10} and
\citet{fang10} confirmed this initial detection and found two new X-ray
absorption systems associated with higher-redshift walls.  
Conversely, \citet{williams10} associated a $\sim 20$\,Mpc galaxy filament
with a previously-detected X-ray absorber.
While the number of detected systems is yet small, the emerging picture 
seems to be that warm-hot gas pervades such large-scale structures.

However, individual galaxies also host warm-hot, extended coronae 
\citep[e.g.][]{anderson10,anderson11,dai12,gupta12,mulchaey10,williams05}, 
as do low-mass galaxy groups \citep{mahdavi00}. As with any absorption-line
study, full spectroscopic information about
virialized structures near the line of sight \citep[e.g.][]{prochaska11} is
critical to our interpretation of the detected lines.
To this end, we are undertaking a survey of galaxies around four quasars
where intervening X-ray absorption systems have been reported; here 
we present initial results from the H2356 sightline.
A concordance $\Lambda$CDM cosmology with $\Omega_m=0.3$, 
$\Omega_\Lambda=0.7$, and $H_0=70$\,km\,s$^{-1}$\,Mpc$^{-1}$.

\section{Data and galaxy samples} \label{sec_data}
\citet{buote09} and \citet{zappacosta10} (hereafter B09 and Z10 respectively) 
analyzed \chandra\ 
and \emph{XMM-Newton} grating observations of H2356; B09 using a 
combined $\sim 230$ ksec in \chandra--LETG and XMM--RGS,
and Z10 using a total of 600\,ksec of \chandra--LETG observations. 
B09 reported the detection of a significant \ovii\ K$\alpha$ absorption
line at the redshift of the $z=0.03$ Sculptor Wall \citep[later confirmed with deeper data by][]{fang10}.
Subsequently, Z10 jointly analyzed these spectra with the
2dF Galaxy Redshift Survey \citep[2dFGRS;][]{colless01}
catalog to search for WHIM near other large-scale structures, detecting low-significance
candidate absorption systems near the Pisces-Cetus Supercluster ($z=0.062$) and the 
Farther Sculptor Wall ($z=0.128$). In a more recent study, \citet{zappacosta12} 
reported yet another X-ray absorption line, C\,{\sc v} at $z=0.112$. Interestingly,
this one appears to lie in a relatively empty region with the nearest galaxies $>2$\,Mpc away,
and its redshift is marginally consistent with a large-scale filament of galaxies.

For our analysis, we extract galaxies in a $140^\prime \times
140^\prime$ region centered on H2356 ($\alpha=23^{\rm h}59^{\rm m}07\fs 9,
\delta=-30^{\rm d}37^{\rm m}41^{\rm s}$) from the 2dFGRS.  This corresponds to
$\sim 5\times 5$\,Mpc at $z=0.03$, the redshift of the nearest absorption
system.  Since the 2dFGRS is incomplete at faint magnitudes and in crowded
regions, we supplemented these public datasets with our own multi-object
spectroscopic survey around this quasar.  For this we employed the
Inamori-Magellan Areal Camera and Spectrograph \citep[IMACS;][]{dressler11} on
the 6.5\,m Magellan-Baade telescope in f/2 mode with a circular, 27$^\prime$
diameter field of view.  Bright galaxies were selected from the NASA/IPAC
Extragalactic Database (NED), while fainter ones very close to the sightline
were taken from our own Magellan imaging. Masks were observed for
at least 1 hour each, employing a 300 lines mm$^{-1}$ grism
covering 3900-10000\AA\ at a dispersion of 1.34 \AA\
pixel$^{-1}$. Spectra were reduced and fit using the techniques employed by
\citet{bai10}. Ultimately we obtained redshifts for 226 galaxies. Between
redshifts from the literature and our own survey, our sample is at least $75$\%
complete to $R2<20$, which corresponds to relatively faint absolute magnitudes
($M_r\sim -15.4, -17.0$, and $-18.7$ at $z=0.030, 0.62$, and $0.128$
respectively).  However, most of these apparently faint galaxies are
at high redshifts ($z>0.2$). Through the combination of 2dF and IMACS spectra,
selected from a variety of imaging data sets, it is unlikely that our sample
is missing faint galaxies near (or bright galaxies farther from) the absorbers.

\section{Galaxies and groups near the absorption systems} \label{sec_filaments}
\subsection{Associating absorbers with halos}
Since galaxies, groups, clusters, and large-scale filaments
and walls are not mutually exclusive classifications, it is often
difficult (or impossible) to uniquely categorize an absorption line
as arising from any one nearby structure (particularly when precise 
velocities are not available, as is the case with X-ray spectroscopy). 
However, it is possible to search for commonalities over an ensemble
of systems; e.g., whether a \emph{specific type} of galaxy or galactic
system is typically found near the warm-hot gas. Broadly speaking,
more massive galaxies and groups will contain more gas at larger
extents, and so their warm-hot gas may be detected at higher impact
parameters than that in low-mass galaxies.
To take this into account, in the following analysis we adopt the host 
dark matter halo's 
virial radius as a rough dividing line for ``associated'' vs.
``unassociated'' systems. Although this line is somewhat arbitrary,
recent theoretical studies suggest that using $R_{vir}$ should be
a reasonable (or even conservative) criterion \citep[e.g.][]{furlanetto05,sharma12}.

\begin{figure}
\plotone{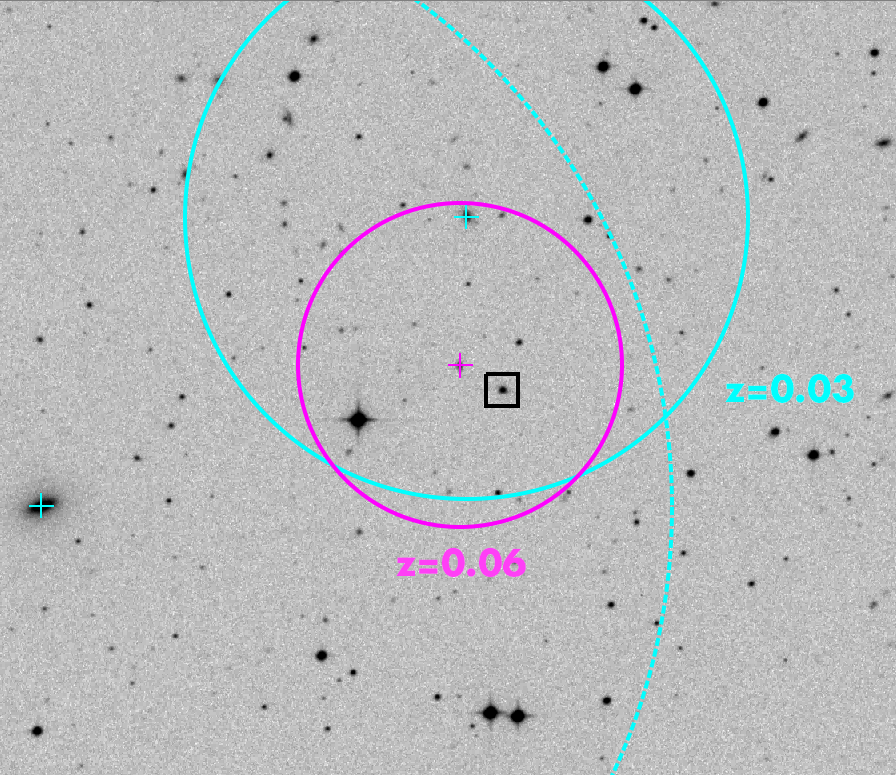}
\caption{$12\farcm5 \times 10\farcm 8$ UKST/SuperCOSMOS red image \citep{hambly01a} 
of the field near H2356. The quasar is marked by the black
square, with solid cyan and magenta circles denoting the virial radii of the nearest
galaxies at $z=0.03$ and $z=0.062$ respectively. Additionally, the large dashed cyan arc
shows the virial radius of the bright $z=0.03$ galaxy to the left, which was included in
the Y05 group catalog and is shown as a circle in Figure~\ref{fig_grpsky}.
\label{fig_image}}
\vspace{0.5cm}
\end{figure}

The virial radii of dark matter halos with \citet{nfw} profiles scale with
mass according to the following simple relation:
\begin{equation}
R_{vir} = \left(\frac{2GM_{vir}}{\Delta_c H_0^2}\right)^{1/3}
\end{equation}
where $\Delta_c\sim 100$ in the concordance $\Lambda$CDM cosmology.
Following \citet{mulchaey09}, for each galaxy in the 2dF sample we estimate $M_{vir}$ 
from the \citet{tinker09} $M_r$--minimum $M_{halo}$ relation. 
SDSS $r$ magnitudes are estimated with the formula 
\begin{equation}
r = R2+0.07(b_J-R2)+0.13
\end{equation}
which is derived from the color transformations given in \citet{blair82}
and \citet{fukugita96}; $b_J$ and $R2$ are photographic
SuperCOSMOS magnitudes.  These magnitudes have large absolute uncertainties
\citep[$\sigma_{R2}\sim 0.3$ at $19<R2<20$;][]{hambly01b}; however, since
$R_{vir}\sim M_{vir}^{1/3}$ these errors typically correspond to just
$\sim 10$\% uncertainties on $R_{vir}$.  For comparison purposes,
we additionally adopt the characteristic $r-$band magnitude
$M_r^\star=-21.21$ derived from SDSS by \citet{blanton03}.

\begin{figure*}
\plotone{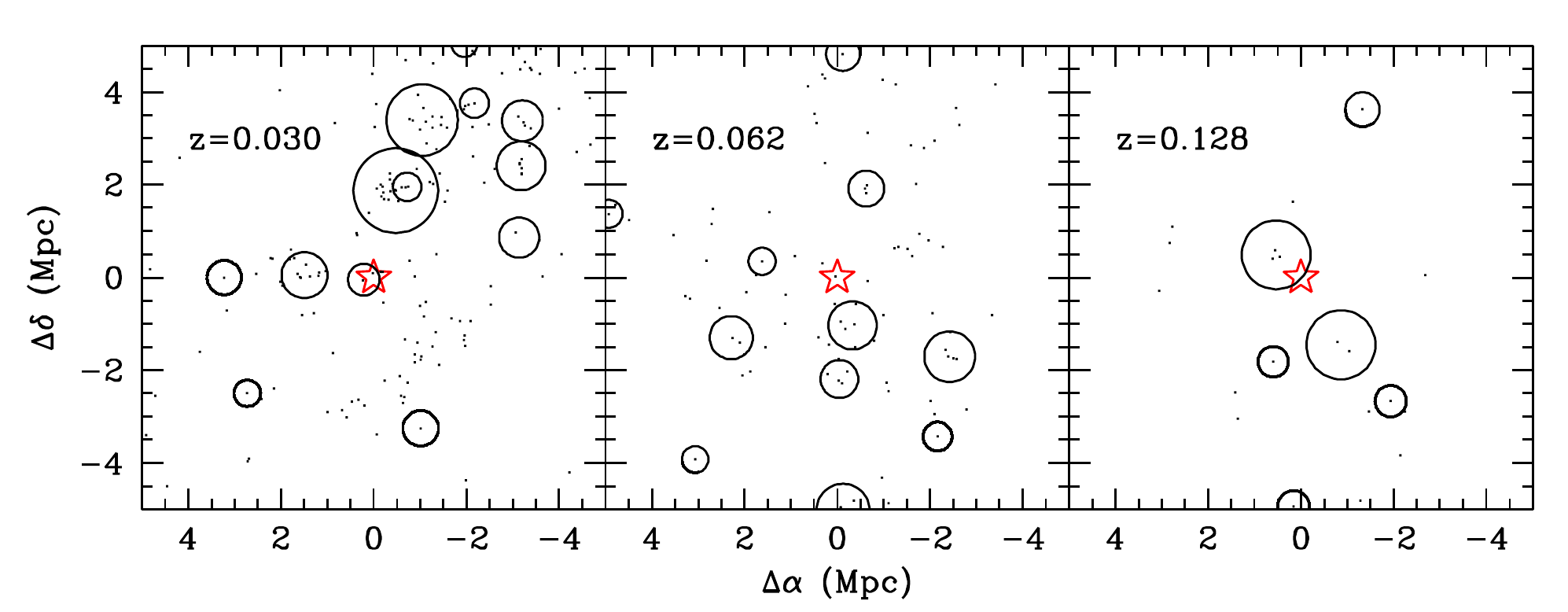}
\caption{$10\times 10$\,Mpc boxes centered on the three absorption systems 
(in $z_{abs}\pm 0.0015$ / $v_{abs}\pm 450$\,km s$^{-1}$ 
redshift slices). Circles show groups from the Y05 2dF halo catalog, where the
radius of each circle indicates the estimated virial radius of each group halo. The lowest
and highest redshift absorbers lie within these projected virial radii. Given this and the individual galaxy
halos shown in Figure~\ref{fig_image}, none of these three absorbers appears to be truly
isolated. \label{fig_grpsky}}
\end{figure*}

A significant fraction of low-redshift galaxies reside
in low-mass, virialized groups, many of which have substantial
reservoirs of warm-hot gas. To determine whether
any of the X-ray absorbers are hosted in group-scale halos,
we incorporate the 2dFGRS group catalog by \citet[][hereafter Y05]{yang05}
into our analysis.  In this catalog, halo masses are assigned using
mass-to-light ratios inferred from the conditional
luminosity function; this catalog is complete to
$M_{halo}\sim 2\times 10^{12}$\,M$_\odot$ at $z=0.12$. 
We use the Y05 halo masses and equation 1 to compute group virial
radii, taking the luminosity-weighted average position of the member 
galaxies in each group as the halo center and redshift. 

\subsection{Sculptor Wall, $z=0.03$}
A strong \ovii\ absorption line at a redshift consistent
with the Sculptor Wall was first reported by B09
and later confirmed by \citet{fang10}, who measured a column
density of $\log (N_{\rm OVII}/{\rm cm}^{-2}) =16.8^{+1.3}_{-0.9}$.
Figure~\ref{fig_image} shows a close-up UKST red image of
the quasar field with the nearest $z=0.030\pm 0.0015$ galaxies
circled in cyan (each galaxy's $R_{vir}$ denoted by the circle's radius).
Although previous studies conclude that the absorption is associated
with WHIM gas permeating the superstructure, in fact the 2dF survey
shows an individual low-luminosity ($0.1L_r^\star$) galaxy at the same redshift,
90 projected kpc to the north. Additionally, a significantly brighter ($1.2L_r^\star$), early-type
galaxy lies 240\,kpc to the east. These galaxies have inferred
$R_{vir}=160$\,kpc and 350\,kpc, respectively, so in principle
the warm-hot gas traced by the \ovii\ could inhabit
either halo (or both).
Note that this more massive galaxy is also listed as a single-member
group in the Y05 catalog (see Figure~\ref{fig_grpsky}). 
Although our IMACS spectroscopy
confirmed the 2dF redshift of the nearer galaxy, no additional nearby galaxies
at this redshift were found.

\subsection{Pisces-Cetus Supercluster, $z=0.062$}
Z10 searched for X-ray absorption systems at the redshift of this 
large-scale filament. While no individual lines were found, by jointly fitting
several marginally-detected ($1-1.5\sigma$) lines and upper limits
they infer the presence of warm-hot ($\log T=5.35^{+0.07}_{-0.13}$) 
gas near the supercluster redshift. Interestingly, they also report tentative
evidence for a separate hotter ($\log T \sim 7$) phase, albeit
at $\sim 1\sigma$ confidence. The 2dFGRS catalog reveals a $0.25L_r^\star$
disk galaxy
at the absorber redshift with impact parameter 49\,kpc and virial
radius 197\,kpc (shown in Figure~\ref{fig_image} as a magenta cross and 
circle); thus, the absorber is again well within the projected virial radius
of the galaxy.  Our IMACS survey reveals several other galaxies within
1\,Mpc, but none are likely to be associated with the absorber (Figure~\ref{fig_imacs}).

\subsection{Farther Sculptor Wall, $z=0.128$}
This absorption system was also detected by Z10 through a joint
fit to the \chandra\ spectrum over several ionic species;
its significance appears to be the most marginal of the three
(with inferred $\log (N_H/{\rm cm}^{-2})=19.8^{+0.4}_{-0.8}$). 
Nonetheless, the Y05 catalog includes a group of three
galaxies 724\,kpc from the absorber, with a halo mass
$2.3\times 10^{13}$\,M$_\odot$ and $R_{vir}=740\,$kpc (Figure~\ref{fig_grpsky}).
Formally the absorber lies within the group's projected virial radius;
however, given the uncertainties involved, 
we can at most conclude that the hot gas at this redshift is marginally
associated with the group halo.

\begin{figure*}
\plottwo{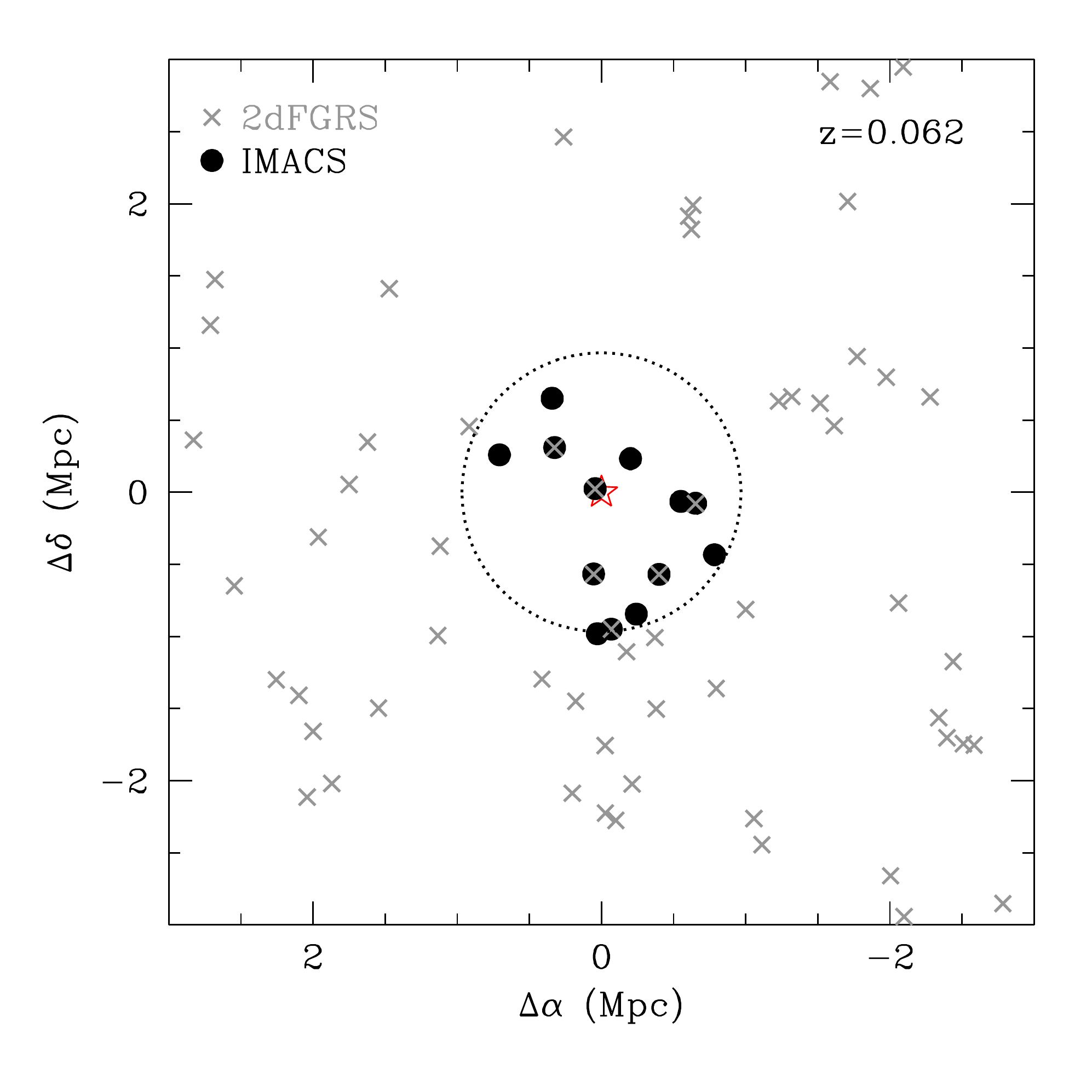}{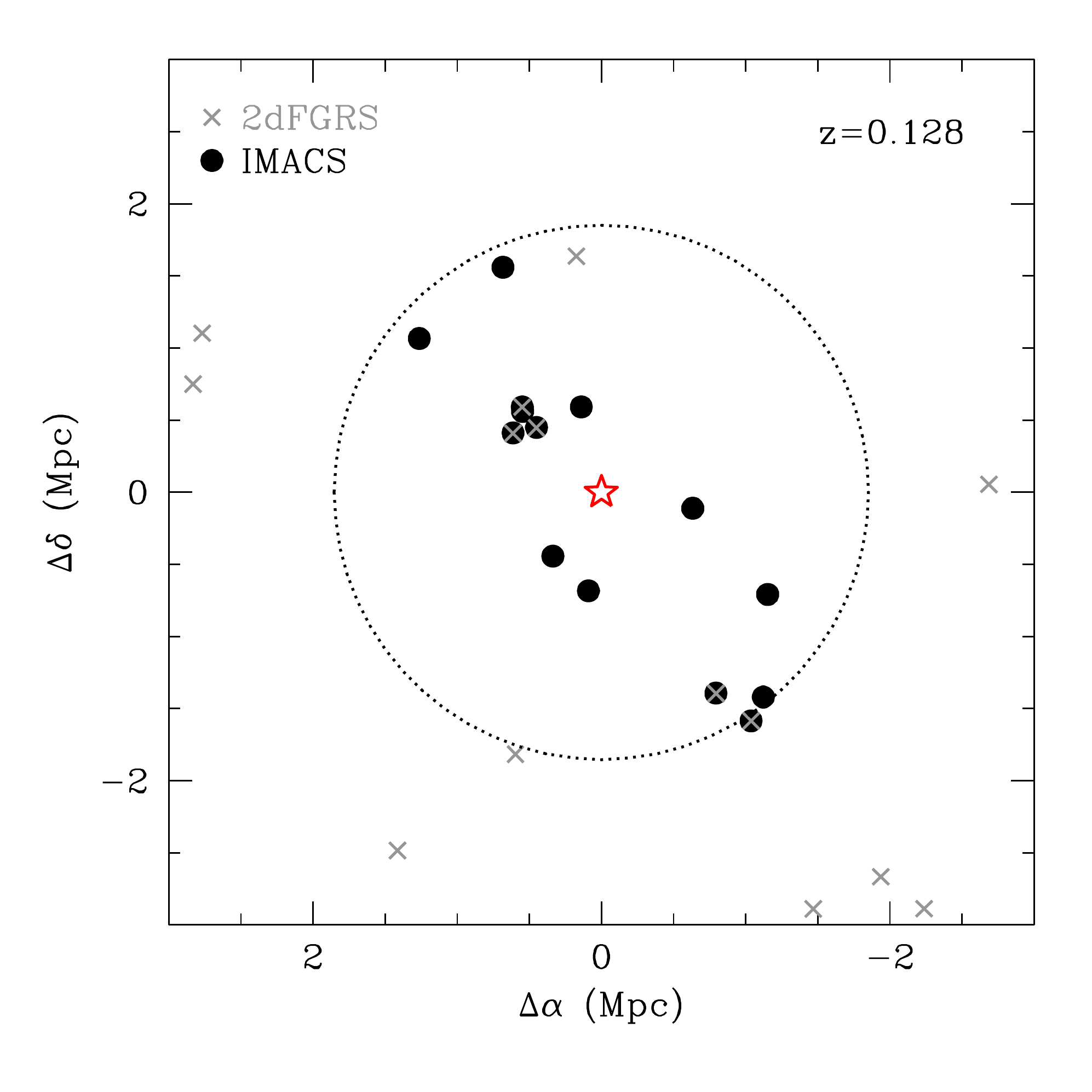}
\caption{Positions of galaxies within $z_{abs}\pm 0.0015$ for the two
higher-redshift absorbers, from our IMACS survey (black circles) and in the
2dFGRS catalog (grey crosses). The dotted circle shows the approximate area of
the IMACS survey. The IMACS spectra confirmed the 2dFGRS redshifts of the
galaxies and groups nearest the absorbers, but no new individual galaxies near
the absorbers were found. If the $z=0.128$ absorber is not associated with the
three-galaxy group to the northeast (despite lying within its projected virial
radius), it may instead be associated with the well-defined galaxy filament
found in our IMACS survey.
\label{fig_imacs}}
\end{figure*}

Our IMACS data show several previously uncatalogued  galaxies within
2\,Mpc of the absorber. None is individually close enough to 
be associated with the hot gas; strikingly, however, they
appear to form a sub-filament within the larger-scale structure
of this wall, nearly centered on the quasar line of sight
(Figure~\ref{fig_imacs}). The warm-hot gas at this redshift may
therefore be associated either with the nearby group, the larger-scale
filament \citep[cf.][]{williams10}, or
a combination of both.

\subsection{A ``void absorber'' at $z=0.112$?}
In their recent reanalysis of the H2356 \chandra\ spectrum,
\citet{zappacosta12} reported a $2.9\sigma$ C\,{\sc v}
absorption line at $z\sim 0.112$; notably, the nearest galaxy
in 2dF at this redshift is about 2.2\,Mpc away. They instead associate
the absorber with a $\sim 30$\,Mpc large-scale filament of galaxies, marginally
consistent in redshift, that passes near the line of sight.
We have checked our IMACS data for galaxies at this redshift
that 2dF may have missed; while two galaxies indeed
fall somewhat closer to the absorber (800 and 1700 projected
kpc), their luminosities are low 
($M_r\sim -20$ and $-18$ respectively). The brightest of the two
has an estimated $R_{vir}\sim 300$\,kpc, much smaller than
the impact parameter to the quasar sightline.
The combined 2dFGRS and IMACS data thus support
the conclusion that this absorption line (if genuine) is not associated
with a massive galaxy or group, though the possibility of an undiscovered faint
nearby galaxy still remains (see Section~\ref{sec_data}). 
Interestingly, \citet{tejos12} also estimate that about 1 in 4
Ly$\alpha$ absorption systems may occur in underdense regions.

\section{Discussion} \label{sec_discussion}
\subsection{The Mrk 421 WHIM-galaxy connection revisited}
In a previous study \citep{williams10}, we examined the galaxy
distributions around two candidate absorption systems detected in the Mrk 421 X-ray spectrum,
finding a large-scale filament coincident with the system at $z=0.027$. The two 
nearest galaxies lie 360 and 390 projected kpc away at the same redshift, a
large enough distance that we discounted the notion of a galactic origin for
that absorption system. However, these galaxies are relatively bright, the
nearer of the two $M_r=-20.9$ -- corresponding to $R_{vir}\sim 500$ kpc.
Moreover, as these two galaxies are separated from each other by only 350 projected kpc and
8\,km\,s$^{-1}$ in radial velocity, they may well share a common group halo,
in which case their virial radius would be even larger.
Thus, under the $R_{vir}$ criterion adopted in this paper, the $z=0.027$ 
Mrk 421 system would in fact be associated with the brighter of these
galaxies (or their group), not the more distant galaxy filament. In the
context of the current study, where all three candidate absorption systems
lie nominally within galaxy or group dark matter halos, a halo origin 
for the Mrk 421 absorber may in fact be more globally consistent.
Interestingly, the nondetection of broad Ly$\alpha$ in this
absorber by \citet{danforth11} may be indicative of higher-metallicity
gas, which would be preferentially found in a bound group halo rather than
a sparse intergalactic filament.

\subsection{The hosts of X-ray absorption line systems}
A fundamental puzzle in any IGM absorption-line study, particularly
one involving metal lines, is whether the systems being measured are connected to
localized phenomena like galaxies (or outflows from the same), or
the more broadly distributed intergalactic medium.  Since metals are strictly
produced in galaxies, \emph{all} metal-line absorption systems must
in some sense have their genesis in stellar systems.  Thus, the issue is really
whether the observed gaseous systems are now \emph{bound} to dark matter halos,
or are instead the free-floating remnants of past outflows which
have over time mixed with pristine intergalactic gas.  
High-redshift Ly$\alpha$ absorbers, even those which likely arise in the IGM,
are often found to be enriched (\citealt{cowie95,ellison00}, but see
\citealt{fumagalli11} for a notable exception), so it stands to reason 
that some metal line systems could serve as signposts for the ``true'' IGM.

However, one unifying theme among galaxy-absorber cross-correlation studies
is that \emph{strong} absorption systems predominantly fall
near galaxies. This appears to hold over a range of redshifts and ionic species,
from damped Ly$\alpha$/Lyman limit systems to O\,{\sc vi} 
\citep[e.g.][]{prochaska11,rudie12,wakker09}. 
Even ions like Ne\,{\sc viii}, which arise primarily in $10^6$\,K 
collisionally-ionized plasma
and were once thought to hold great promise for tracing the WHIM \citep[e.g.][]{savage05}
in fact appear to be located in or near galaxies as well \citep{mulchaey09,narayanan11}
Due to its tenuous and highly-ionized nature, the intergalactic medium 
seems to make its presence known mostly (or exclusively) in low column 
density systems (like Ly$\alpha$ forest lines; $N_{\rm HI}\sim 10^{14}$\,cm$^{-2}$). 

In this context, it would not be surprising if all currently known X-ray
absorption line systems originate in bound systems. State-of-the-art
X-ray spectrographs like \chandra\ and \emph{XMM-Newton}, while revolutionary,
are ultimately limited by the tyranny of Poisson statistics and the faintness
of background AGN.  Even the most exquisite grating spectra achieve 
sensitivities no better than $N_{\rm OVII}\sim 10^{15}$\,cm$^{-2}$, or an equivalent
hydrogen column density of at least $N_H\sim 10^{19}$\,cm$^{-2}$; typical deep
spectra are several times less sensitive. Thus, any 
significant X-ray line likely represents a relatively dense, enriched system.
While a fraction of these regions may exist as filaments in intergalactic space, 
they are almost certainly ubiquitous in the form of circumgalactic and intragroup media.

The galactic structures surrounding X-ray absorption systems, whether individual galaxies,
groups, voids, or filaments, therefore play a key role in interpreting
the nature of those systems.
All told, four X-ray absorbers that were initially presented as evidence for
the WHIM (that is, the three from \citealt{buote09} and \citealt{zappacosta10},
along with the higher-redshift Mrk 421 absorber) may instead be associated
with bound dark matter halos; two other systems ($z=0.011$ toward Mrk 421;
\citealt{nicastro05a}, and the $z=0.112$ system of \citealt{zappacosta12}) 
currently have no known galaxies or groups nearby. It cannot currently determined
whether these absorbers
originate in the WHIM, very faint nearby galaxies, or are themselves spurious detections.
However, the emerging picture -- admittedly 
limited by small-number statistics --  suggests that a
significant fraction of X-ray absorption lines are hosted in bound systems.
This remaining ambiguity illustrates the need for not just deep X-ray spectroscopy
of multiple background quasars,
but also complete galaxy surveys around the quasar sightlines to fully assess
the nature of the surrounding structure. Our Magellan/IMACS survey is ongoing;
an analysis of the full galaxy sample with the most recent \chandra\ spectral data
along multiple sightlines will be presented in a forthcoming paper.

Unfortunately, even with a full sample of X-ray spectra and galaxy redshifts,
systematic uncertainties in the \chandra/LETG wavelength scale prevent us from
kinematically associating absorbers with individual galaxies; in fact no new
instruments with this capability are currently on the horizon.
However, observations and theory suggest that warm-hot absorbers likely occur in multiphase
media \citep{furlanetto05,fox11}. If this is the case, deep high-resolution UV spectroscopy of 
lower-ionization lines
(e.g.~C\,{\sc iv}, O\,{\sc vi}, Ne\,{\sc viii}, and broad Ly$\alpha$) may reveal extended warm gaseous
phases associated with the same galaxies, thereby providing a circumstantial (if not
conclusive) link between the galaxies and the X-ray absorbers.

\section{Summary} \label{sec_summary}
By combining wide-field survey data from 2dF with new Magellan
multi-object galaxy spectroscopy, we have presented a detailed view of the
galaxy and group halo distribution around three low-redshift
X-ray absorption line 
systems toward H2356-309. Each of these systems appears to lie within
the projected virial radius of at least one galaxy and/or group halo at the same
redshift, though for the most distant system the overlap is marginal.
Though our sample is small, we conclude that a significant
fraction of X-ray absorption systems may arise in bound structures,
indicating that some of them may trace extended
circumgalactic (and/or intragroup) media rather than the filamentary warm-hot IGM.

\acknowledgments
We thank the anonymous referee for suggestions which improved the manuscript.
Support for this work was provided by the National Aeronautics and Space
Administration through Chandra Award Number AR2-13016X issued by the Chandra X-ray
Observatory Center, which is operated by the Smithsonian Astrophysical
Observatory for and on behalf of the National Aeronautics Space Administration
under contract NAS8-03060.  Partial support was also provided by NSF grant AST-0707417.
This research has made use of the NASA/IPAC Extragalactic Database (NED) which is
operated by the Jet Propulsion Laboratory, California Institute of Technology,
under contract with the National Aeronautics and Space Administration; 
and of the SuperCOSMOS Science
Archive, prepared and hosted by the Wide Field Astronomy Unit, Institute for
Astronomy, University of Edinburgh, which is funded by the UK Science and
Technology Facilities Council.


\begin{thebibliography}

\bibitem[Anderson \& Bregman(2010)]{anderson10} Anderson, M.~E.~\& Bregman, 
	J.~N.~2010, \apj, 714, 320
\bibitem[Anderson \& Bregman(2011)]{anderson11} Anderson, M.~E.~\& Bregman,
	J.~N.~2011, \apj, 737, 22
\bibitem[Bai et al.(2010)]{bai10} Bai, L., Rasmussen, J., Mulchaey, J.~S.,
         et al.~2010, \apj, 713, 637
\bibitem[Blair \& Gilmore(1982)]{blair82} Blair, M., \& Gilmore, G.~1982,
         \pasp, 94, 742
\bibitem[Blanton et al.(2003)]{blanton03} Blanton, M.~R., Hogg, D.~W., Bahcall,
         N.~A., et al.~2003, \apj, 592, 819
\bibitem[Buote et al.(2009)]{buote09} Buote, D., Zappacosta, L., Fang, T.,
         et al.~2009, \apj, 695, 1351 
\bibitem[Cen \& Ostriker(1999)]{cen99} Cen, R.~\& Ostriker, J.~P.~1999,
	\apj, 514, 1
\bibitem[Colless et al.(2001)]{colless01} Colless, M., Dalton, G., Maddox, S., 
         et al.~2001, \mnras, 328, 1039
\bibitem[Cowie et al.(1995)]{cowie95} Cowie, L.~L., Songaila, A., Kim, T.-S.,
	\& Hu, E.~M.~1995, \aj, 109, 1522
\bibitem[Dai et al.(2012)]{dai12} Dai, X., Anderson, M.~E., Bregman, J.~N.,
	\& Miller, J.~M.~2012, \apj, 755, 107
\bibitem[Danforth et al.(2011)]{danforth11} Danforth, C.~W., Stocke, J.~T.,
	Keeney, B.~A., et al.~2011, \apj, 743, 18
\bibitem[Dav\'e et al.(2010)]{dave10} Dav\'e, R., Finlator, K., Oppenheimer,
	B.~D., et al.~2010, \mnras, 408, 2051
\bibitem[Dressler et al.(2011)]{dressler11} Dressler, A., Bigelow, B., Hare, T.,
         et al.~2011, \pasp, 123, 288
\bibitem[Ellison et al.(2000)]{ellison00} Ellison, S.~L., Songaila, A., Schaye,
	J., \& Pettini, M.~2000, \aj, 120, 1175
\bibitem[Fang et al.(2002)]{fang02} Fang, T., Marshall, H.~L., Lee, J.~C.,
	Davis, D.~S., \& Canizares, C.~R.~2002, \apj, 572, L127
\bibitem[Fang et al.(2007)]{fang07} Fang, T., Canizares, C.~R., \& Yao, 
	Y.~2007, \apj, 670, 992
\bibitem[Fang et al.(2010)]{fang10} Fang, T., Buote, D.~A., Humphrey, P.~J.,
	et al.~2010, \apj, 714, 1715
\bibitem[Fox(2011)]{fox11} Fox, A.~J.~2011, \apj, 730, 58
\bibitem[Fukugita et al.(1996)]{fukugita96} Fukugita, M., Ichikawa, T., Gunn, J.~E.,
         et al.~1996, \aj, 111, 1748
\bibitem[Fumagalli et al.(2011)]{fumagalli11} Fumagalli, M., O'Meara, J.~M., \&
	Prochaska, J.~X.~2011, Science, 334, 1245
\bibitem[Furlanetto et al.(2005)]{furlanetto05} Furlanetto, S.~R., Phillips, L.~A., \&
         Kamionkowski, M.~2005, \mnras, 359, 295
\bibitem[Gupta et al.(2012)]{gupta12} Gupta, A., Mathur, S., Krongold, Y.,
	Nicastro, F., \& Galeazzi, M.~2012, \apj, 756, L8
\bibitem[Hambly et al.(2001a)]{hambly01a} Hambly, N.~C., MacGillivray, H.~T.,
	Read, M.~A., et al.~2001a, \mnras, 326, 1279
\bibitem[Hambly et al.(2001b)]{hambly01b} Hambly, N.~C., Irwin, M.~J., \&
         MacGillivray, H.~T.~2001b, \mnras, 326, 1295
\bibitem[Mahdavi et al.(2000)]{mahdavi00} Mahdavi, A., B\"ohringer, H.,
	Geller, M.~J., \& Ramella, M.~2000, \apj, 534, 114
\bibitem[Mulchaey \& Chen(2009)]{mulchaey09} Mulchaey, J.~S.~\& Chen, H.-W.~2009,
	\apj, 698, L46
\bibitem[Mulchaey \& Jeltema(2010)]{mulchaey10} Mulchaey, J.~S.~\& Jeltema, T.~E.~2010,
	\apj, 715, L1
\bibitem[Narayanan et al.(2011)]{narayanan11} Narayanan, A., Savage, B.~D., Wakker, B.~P.,
	et al.~2011, \apj, 730, 15
\bibitem[Navarro, Frenk, \& White(1996)]{nfw} Navarro, J.~F., Frenk, C.~S., \& White,
	S.~D.~M.~1996, \apj, 462, 563
\bibitem[Nicastro et al.(2005a)]{nicastro05a} Nicastro, F., Mathur, S., Elvis, M.,
         et al.~2005a, \apj, 629, 700 
\bibitem[Nicastro et al.(2005b)]{nicastro05b} Nicastro, F., Mathur, S., Elvis, M.,
         et al.~2005b, Nature, 433, 495
\bibitem[Perna \& Loeb(1998)]{perna98} Perna, R.~\& Loeb, A.~1998, \apj, 503, 135L
\bibitem[Prochaska et al.(2011)]{prochaska11} Prochaska, J.~X., Weiner, B., Chen,
	H.-W., Mulchaey, J., \& Cooksey, K.~2011, 740, 91
\bibitem[Rudie et al.(2012)]{rudie12} Rudie, G.~C., Steidel, C.~C., Trainor, R.~F.,
	et al.~2012, \apj, 750, 67
\bibitem[Savage et al.(2005)]{savage05} Savage, B.~D., Lehner, N., Wakker, B.~P.,
	Sembach, K.~R., \& Tripp, T.~M.~2005, \apj, 626, 776
\bibitem[Sharma et al.(2012)]{sharma12} Sharma, P., McCourt, M., Parrish, I.~J.,
         \& Quataert, E.~2012, MNRAS, in press (arXiv:1206.4314)
\bibitem[Tejos et al.(2012)]{tejos12} Tejos, N., Morris, S.~L., Crighton, N.~H.~M.,
         et al.~2012, \mnras, 425, 245
\bibitem[Tinker \& Conroy(2009)]{tinker09} Tinker, J.~L., \& Conroy, C.~2009,
	\apj, 691, 633
\bibitem[Wakker \& Savage(2009)]{wakker09} Wakker, B.~P.~\& Savage, B.~D.~2009,
	\apjs, 182, 378
\bibitem[Williams et al.(2005)]{williams05} Williams, R.~J., Mathur, S., 
        Nicastro, F., et al.~2005,
	\apj, 631, 856
\bibitem[Williams et al.(2007)]{williams07} Williams, R.~J., Mathur, S.,
	Nicastro, F., \& Elvis, M.~2007, \apj, 665, 247
\bibitem[Williams et al.(2010)]{williams10} Williams, R.~J., Mulchaey, J.~S.,
	Kollmeier, J.~A., \& Cox, T.~J.~2010, \apj, 724, L25
\bibitem[Yang et al.(2005)]{yang05} Yang, X., Mo, H.~J., van den Bosch, F.~C.,
	\& Jing, Y.~P.~2005, \mnras, 356, 1293 (Y05)
\bibitem[Zappacosta et al.(2010)]{zappacosta10} Zappacosta, L., et al.~2010,
	\apj, 717, 74 (Z10)
\bibitem[Zappacosta et al.(2012)]{zappacosta12} Zappacosta, L., Nicastro, F.,
	Krongold, Y., \& Maiolino, R.~2012, \apj, 753, 137
\end{thebibliography}
\end{document}